\documentclass[12pt]{article}
\usepackage{graphicx,amsmath}
\usepackage{units}
\usepackage{color}

\newlength{\dinwidth}
\newlength{\dinmargin}
\renewcommand{\vec}[1]{\boldsymbol{#1}}

\def\lapproxeq{\lower .7ex\hbox{$\;\stackrel{\textstyle                                                    
<}{\sim}\;$}}                                                    
\def\gapproxeq{\lower .7ex\hbox{$\;\stackrel{\textstyle                                                    
>}{\sim}\;$}}                                                    
\def\be{\begin{equation}}                                                    
\def\ee{\end{equation}}                                                    
\def\bea{\begin{eqnarray}}                                                    
\def\eea{\end{eqnarray}}

\def\GeV{\rm GeV}

\def\sh{\hat s}
\def\sh2{{\hat s}^2}

\begin{document}

\begin{flushright}                                                    
IPPP/14/86  \\
DCPT/14/172 \\                                                    
\today \\                                                    
\end{flushright} 

\vspace*{0.5cm}

\begin{center}
{\Large \bf $t$ dependence of the slope }\\
\vspace*{0.5cm}
{\Large \bf of   the high energy elastic $pp$ cross section.}\\

\vspace*{1cm}
                                                   
V.A. Khoze$^{a,b}$, A.D. Martin$^a$ and M.G. Ryskin$^{a,b}$  \\                                                    
                                                   
\vspace*{0.5cm}                                                    
$^a$ Institute for Particle Physics Phenomenology, University of Durham, Durham, DH1 3LE \\                                                   
$^b$ Petersburg Nuclear Physics Institute, NRC Kurchatov Institute, Gatchina, St.~Petersburg, 188300, Russia \\          
                                                    
\vspace*{1cm}

\begin{abstract} 

We consider the main factors which cause the variation of the value of the local
slope of the elastic $pp$ cross section $B(t)=d[\ln(d\sigma_{\rm el}(pp)/dt]/dt$ with $t$. Namely, we discuss the role of the pion-loop insertion in the pomeron trajectory, the $t$-dependence of the pomeron-nucleon coupling and the role of the eikonalization of the proton-proton amplitude in both the  one- and two-channel eikonal models.
\end{abstract}                                                        
\vspace*{0.5cm}                                                    
                                                    
\end{center}

\section{Introduction  \label{sec:1}} 
The simplest approximation for the $t$-behaviour of the high energy proton-proton differential elastic cross section is to assume that it is described by an exponent, 
\be
d\sigma_{\rm el}/dt=d\sigma_{\rm el}/dt\big|_{t=0}\cdot\exp(Bt),
\ee
where $t$ is the square of the four-momentum transfer, and $B$ is called the $t$-slope. The value of $B$ increases with energy. However more precise data indicate that actually the elastic cross section has a more complicated $t$-behaviour. In particular, the recent
TOTEM data~\cite{Totem} demonstrate that the local slope 
\be
B=d[\ln(d\sigma_{\rm el}/dt]/dt
\ee
 varies with $t$. It is important to understand these variations, for example, when we extrapolate the data to $t=0$ in order to determine the value of total $pp$ cross section based on the optical theorem. The question was considered in~\cite{C18} 14 years ago\footnote{See also \cite{Dremin} for a recent review.}. Here we update the discussion of these variations based on the new LHC data and on the improved understanding of high energy diffractive processes.

The $|t|$ dependence of the local slope $B$ has several different components. In the present paper we use the formalism of Regge theory to discuss the main factors which affect the value of the local slope $B$ considering, in particular, the region of small $t$. 
We shall proceed step-by-step so as to expose the influence of the different components. 
In Sect.2 we consider a simplified model in which the proton-proton amplitude is described by single pomeron exchange. In this case the $t$ dependence of $B$ may be caused by the non-linearity of the pomeron trajectory or by a non-exponential $t$ dependence of the proton-pomeron coupling. In general, we have no reason to expect a linear form of the pomeron trajectory, $\alpha_P(t)=\alpha_P(0)+\alpha'_Pt$. This is an approximation. The nearest singularity at $t=4m^2_\pi\simeq 0.08$ GeV$^2$ corresponds to the production (in $t$-channel) of a pair of pions, and this threshold leads to a non-linear dependence of $\alpha_P(t)$ on $t$. Tuning the parameters in order to reproduce the experimental behaviour of the differential elastic cross section 
(at least in the relatively low $|t|$ domain) we demonstrate numerically the possible role of this pion-loop insertion in the (linear in the first approximation) pomeron trajectory. Then in the next subsection we evaluate the effect caused by replacing the exponential pomeron-proton coupling by the electromagnetic proton form factor $F_1(t)$, as, for example, used in the Donnachie-Landshoff parametrization~\cite{DL1}. Both effects considered in this section lead to a decrease of $B$ with increasing $|t|$.

In Sect.3 we account for the `eikonal' rescattering generated by two-particle $s$-channel unitarity. At low $|t|$, before the first diffractive minimum, these absorptive corrections lead to a growth of the local slope $B$ with $|t|$. In order to be in agreement with the data for low-mass proton diffractive dissociation  we consider not only the one-channel, but also the two-channel eikonal model. We use the Good-Walker formalism and show the expected $t$-dependence of $B$ for different collider energies. The final predictions for the $t$-dependence of $B$ at various collider energies are shown in Fig.8.  Sect.4 contains a discussion and a comparison of the predictions of our model for $B(t)$ with preliminary TOTEM data at 8 TeV, as well as with data at CERN S$p\bar{p}$S and Tevatron energies.

\section{One-pomeron exchange}
The elastic proton-proton amplitude given by the one-pomeron exchange reads
\be
\label{one}
A(s,t)=\sigma_0s_0\beta_N^2(t)\left(\frac s{s_0}\right)^{\alpha_P(t)}[i+\tan(\pi(\alpha_P(t)-1)/2)]\ ,
\ee
where the expression in square brackets, $[...]$, is the signature factor. Following convention, we take the dimensionful scale $s_0=1$ GeV$^2$. 
We use a normalization such that the differential elastic cross section is given by
\be
\label{norm}
\frac{d\sigma_{\rm el}}{dt}~=~\frac{|A(s,t)|^2}{16\pi s^2}\ .
\ee

We first consider the simplest case with a linear pomeron trajectory
\be
\label{traj-l}
\alpha_P(t)~=~1+\Delta+\alpha'_Pt
\ee
and a pure exponential form of the proton-pomeron coupling,
$\sqrt{\sigma_0}\beta(t)$; that is $\beta(t)=\exp(b_{\rm exp}t)$.

We tune the values of the parameters so as  to describe the data in the low $|t|\lapproxeq 0.3$ GeV$^2$ domain. Explicitly, we find $\Delta=0.08,\ \sigma_0=23$ mb, $b_{\rm exp}=1.5$ GeV$^{-2}$ and $\alpha'_P=0.37$ GeV$^{-2}$. This gives a local $B$ slope which increases with energy (due to $\alpha'_P$), but which does not depend on $t$; see\footnote{Note that in Figs.1-7 we are only attempting to describe data in the small $t$ domain, $|t| \lapproxeq 0.3~\GeV^2$. Not until Fig.8 is the model tuned to describe data in a larger $|t|$ domain.} Fig.1.

\begin{figure} 
\begin{center}
\vspace{-3.5cm}
\includegraphics[height=11.5cm]{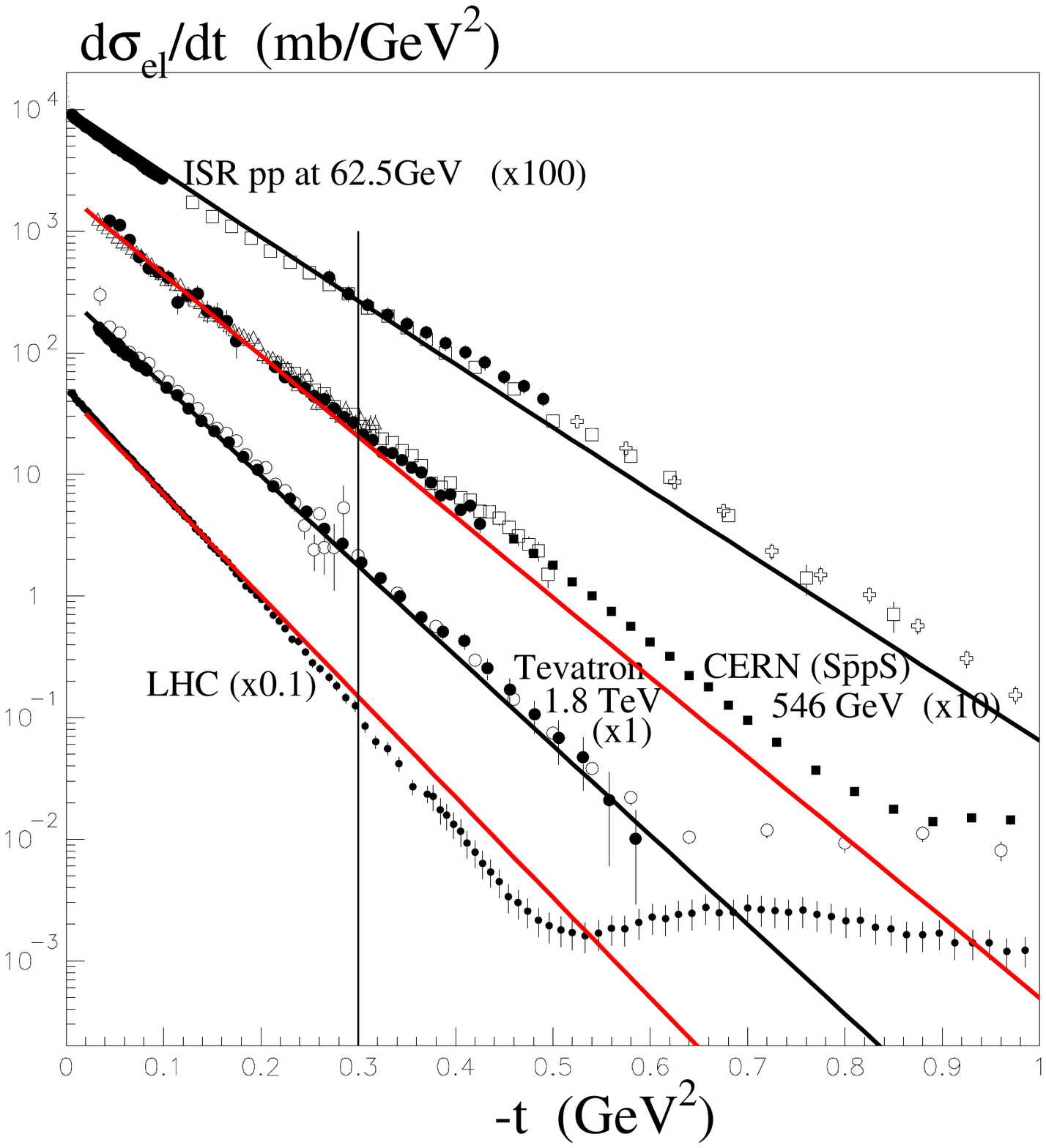}
\includegraphics[height=11.5cm]{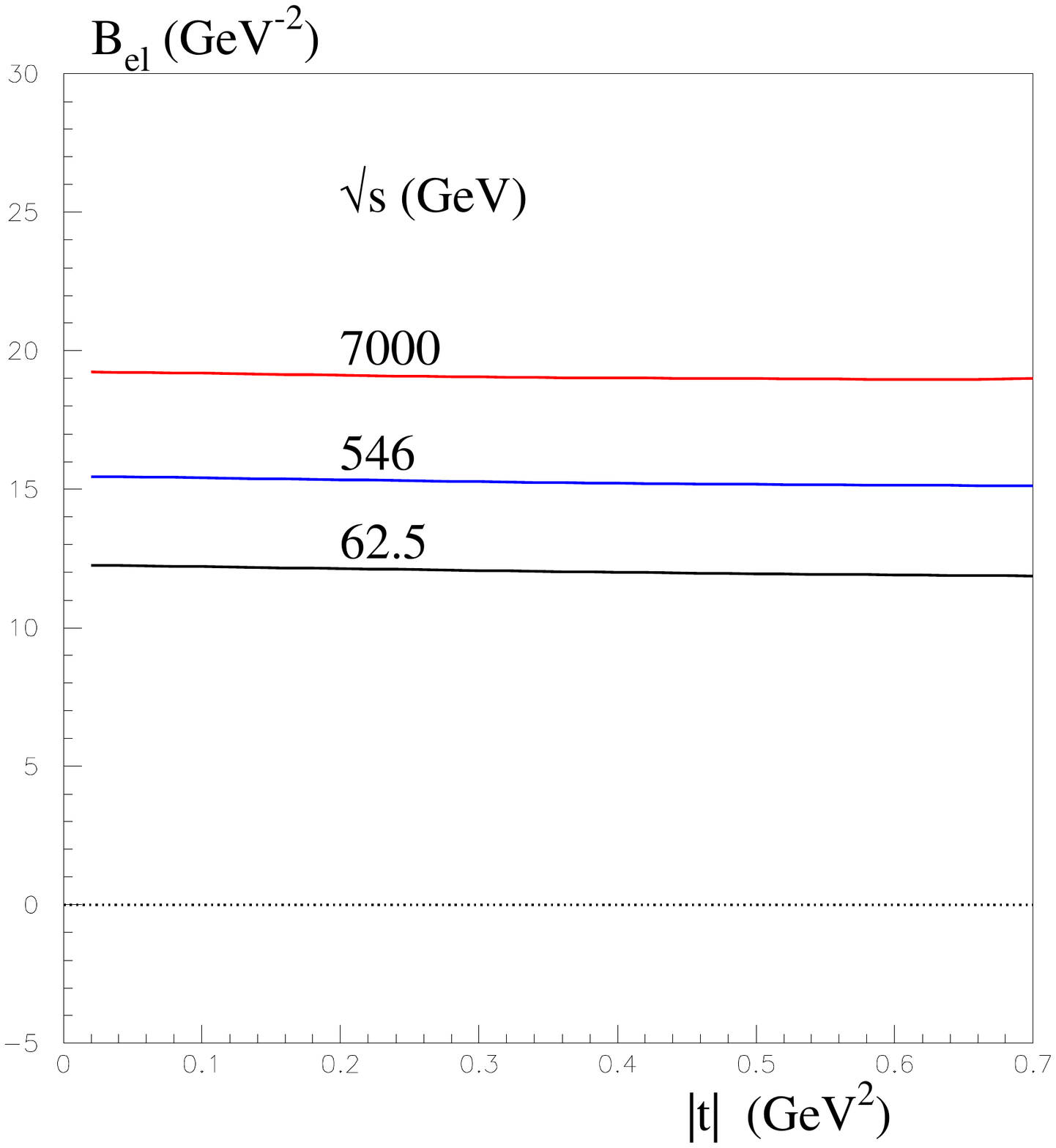}
\caption{\sf The differential proton-proton elastic cross section (left) and the local slope $B$ (right) described by a one pomeron amplitude with a pure exponential coupling and a linear pomeron trajectory; the data are from~\cite{ISR}-\cite{TOTEM}.  The vertical line at $-t=0.3~\GeV^2$ is simply for ease of reference.} 
\label{fig:1}
\end{center}
\end{figure}
Note that the `effective' value of $\alpha'_P$, observed experimentally as the shrinkage of the diffractive cone, increases with energy (see, e.g.~\cite{SR}). Thus, in order to be in approximate agreement with the LHC elastic data,  in the present one pomeron model we have had to use a value of $\alpha'_P=0.37$ GeV$^{-2}$ larger than the canonical value $\alpha'_P=0.25$ GeV$^{-2}$ of the Donnachie-Landshoff~\cite{DL1} fit. In a more complicated eikonal model (considered in the next section), the increase of $\alpha'_{\rm eff}$ is provided by the screening effect described by multi-pomeron diagrams. The absorptive corrections suppress the contribution in centre of the disk and in this way enlarge the effective interaction radius.

\subsection{Insertion of the $\pi$-loop}
Now let us account for the nearest $t$-channel singularity in the pomeron trajectory. It is given by the inclusion of a pion loop.  As was shown in ~\cite{AG},  the trajectory now takes the non-linear form
\be
\label{pi}
\alpha_P(t)~=~1+\Delta+\alpha'_Pt+\frac{\sigma_0(\pi\pi)m^2_\pi}{32\pi^3}~
\beta^2_\pi(t)~h(4m^2_\pi/|t|)\, ,
\ee
where the final factor in the new term
\be
\label{h-pi}
h(\tau)=-\frac{4\beta^2_\pi(t)}{\tau}\left[2\tau-(1+\tau)^{3/2}\ln\left(
\frac{\sqrt{1+\tau}+1}{\sqrt{1+\tau}-1}\right)+\ln\frac{m^2}{m^2_\pi}\right]
\ee
with $\tau=4m^2_\pi/|t|$ and $m=1$ GeV. The factor $\sigma_0(\pi\pi)$ in (\ref{pi}) specifies the
value of pion-pomeron coupling. For this we use the additive quark model result 
\be
\sigma_0(\pi\pi)=(2/3)^2\sigma_0=(4/9)\sigma_0(pp).
\label{eq:8}
\ee
 The factor $\beta_\pi(t)$ accounts for the $t$-dependence of this coupling, for which we take the pole expression 
$\beta_\pi=1/(1-t/b_\pi)$ with $b_\pi=m^2_\rho$.

The cross sections and the local slope $B$ obtained in this case are shown in Fig.2. Some curvature in the $t$-dependence of the local slope is evident in the low $|t|<0.1$ GeV$^2$ region, especially at the larger energies. The effect is rather weak, due to the small numerical value of the factor $\sigma_0(\pi\pi)m^2_\pi/32\pi^3\sim 0.5\times 10^{-3}$  
in the last term of (\ref{pi}). The parameters turn out to be practically the same as before, but with a bit smaller value of $\alpha'_P$; namely 
$\Delta=0.08,\ \sigma_0=23$ mb, $b_{\rm exp}=1.5$ GeV$^{-2}$ and $\alpha'_P=0.36$ GeV$^{-2}$.

It is appropriate to ask how much flexibility exists in the pion-loop contribution. First, we discuss $\beta_{\pi}(t)$.
The pole expression form for $\beta_\pi$
is the standard choice. However, the results do not noticeably depend on the explicit
form of the $t$-dependence provided that this dependence reproduces a
reasonable value of the mean $t$-slope.  Actually, the main assumption, concerning the value of the pion-loop contribution, is in (\ref{eq:8}), where
the additive quark model result (that is the factor $(2/3)^2$) was used. On the other
hand, this ratio $\sigma(\pi p)/\sigma(pp)\simeq 2/3$ is confirmed by the
data at lower energies. We can estimate the possible size of the effect looking at Fig.9
below, where we present the (dashed) curves calculated without the pion loop contribution
 (that is by replacing the factor (2/3) by 0). This {\em extreme}
example leads to about a 1.5 GeV$^{-2}$
variation of the slope within the $0<-t<0.15$ GeV$^2$
interval, which corresponds to a value of the coefficient $c\sim 5$ GeV$^{-4}$ in the
parametrization $d\sigma_{\rm el}/dt=N\exp(bt+ct^2)$. Thus, making the conservative assumption that the
pion-pomeron coupling (the factor 2/3) is known with 25\% accuracy,
we expect less than a 0.05\% deviation in the extrapolation of the differential cross section from $t=-0.02$
GeV$^2$ to $t=0$, coming from the uncertainty in the pion-loop contribution.

\begin{figure} 
\begin{center}
\vspace{-3.5cm}
\includegraphics[height=11.5cm]{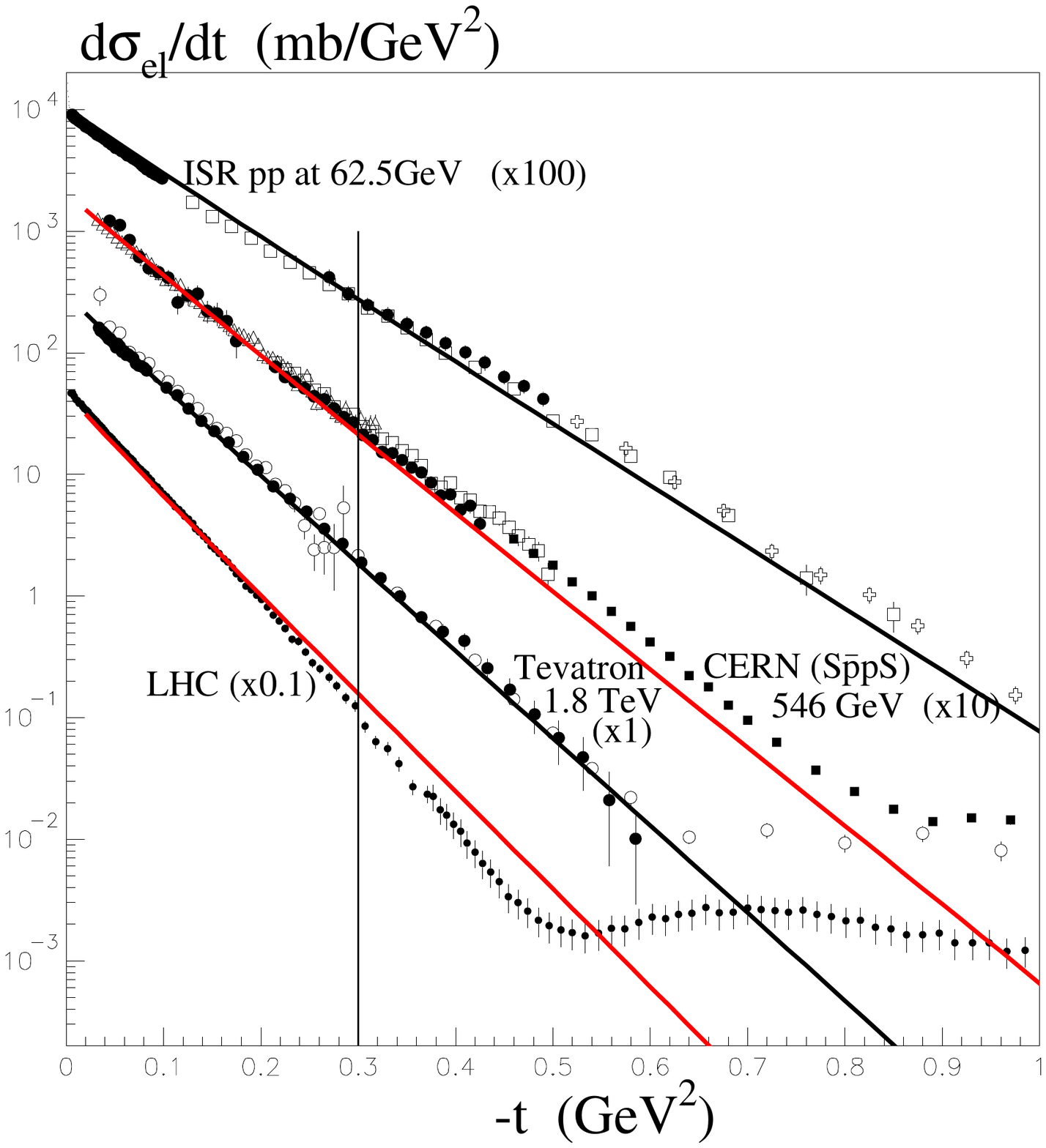}
\includegraphics[height=11.5cm]{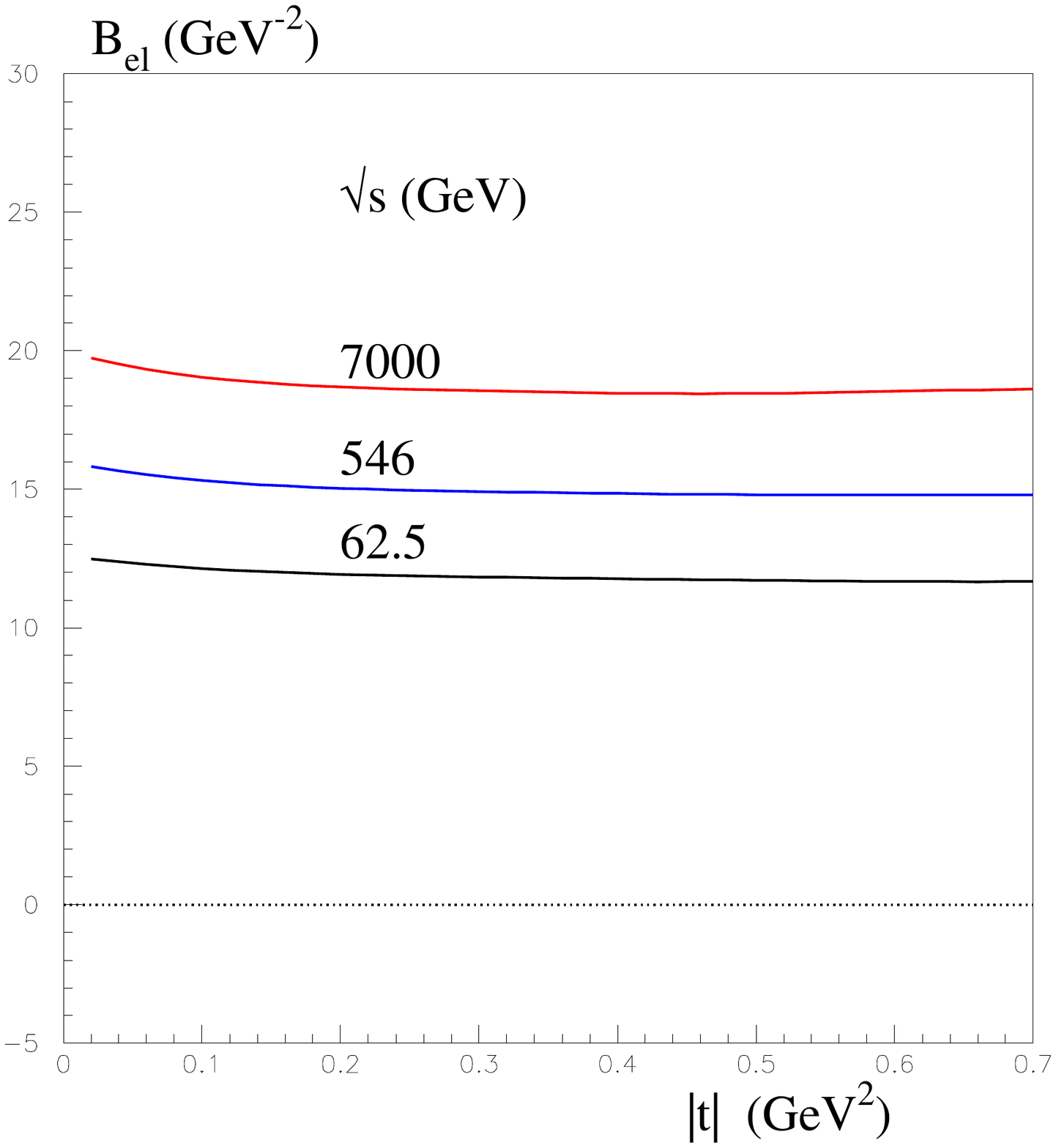}
\caption{\sf The differential proton-proton elastic cross section (left) and the local slope $B$ (right) described by a one pomeron amplitude with a pure exponential coupling and a pion loop included in the pomeron trajectory; the data are from~\cite{ISR}-\cite{TOTEM}.} 
\label{fig:2}
\end{center}
\end{figure}

\subsection{Non-exponential coupling}
Another possible source of the curvature, or $t$-dependence of local slope $B(t)$, is the non-exponential form of the proton-pomeron coupling. Strictly speaking, there is no reason for a pure exponent. It is often used just for convenience. Another popular idea is to assume that this coupling looks like the electromagnetic form factor, $F_1(t)$, of proton~\cite{DL1}.
\be
\label{F1}
\beta(t)=F_1(t)=\frac{4m^2_N-\mu_pt}{4m^2_N-t}\cdot\frac 1{(1-t/b_N)^2}
\ee 
with the proton magnetic moment $\mu_p=2.79$ and $b_N=0.71$ GeV$^2$; $m_N$ is the mass of nucleon.

Using a coupling of the form of (\ref{F1}) we find the results presented in Fig.3 (if the pion-loop is not inserted in the pomeron trajectory) and those in Fig.4 if the pion loop is included. The effect of non-exponential coupling is stronger than that of the pion loop. It reveals itself across the whole $t$ interval, but this effect does not depend on energy.
\begin{figure} 
\begin{center}
\vspace{-3.5cm}
\includegraphics[height=11.5cm]{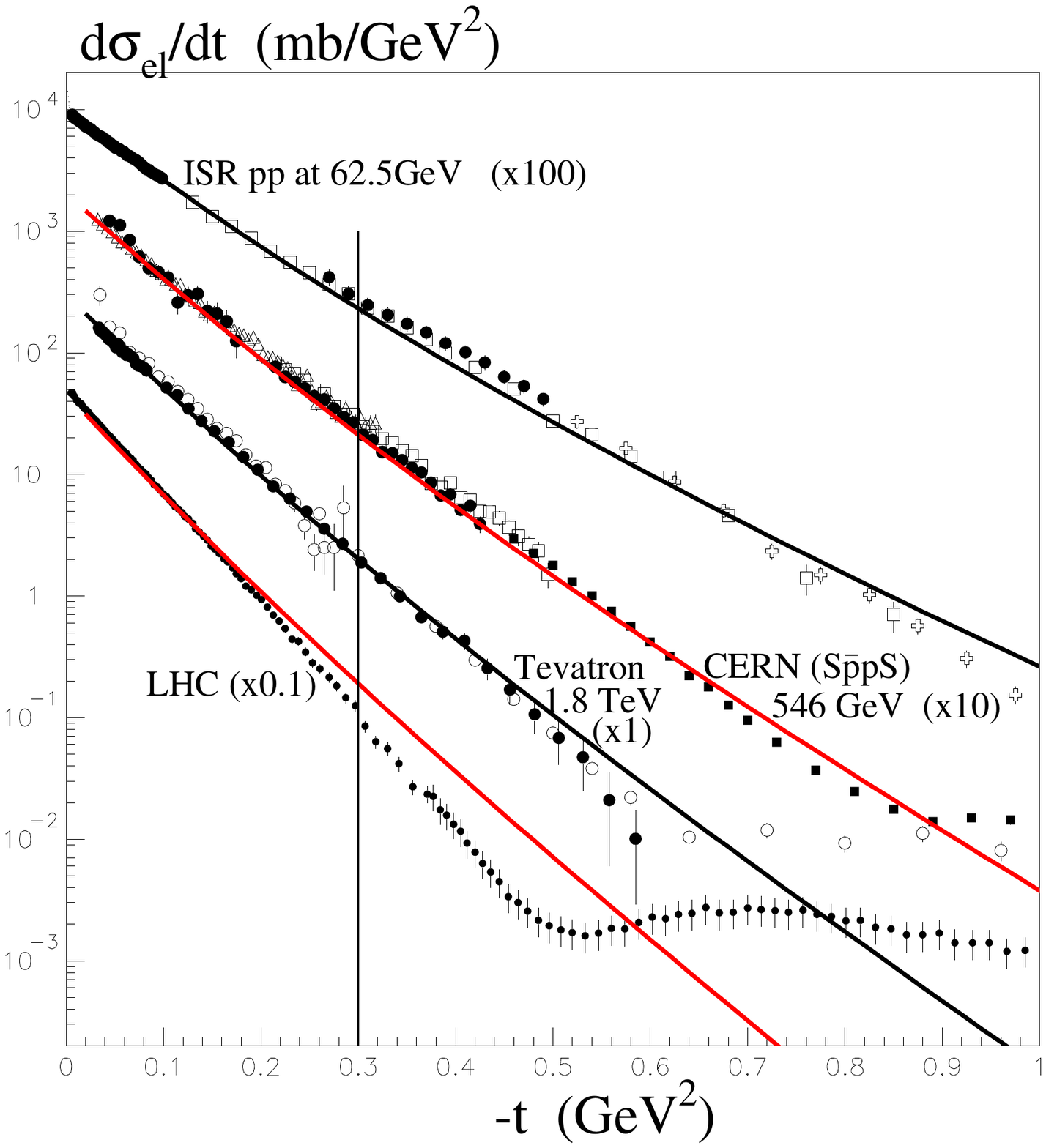}
\includegraphics[height=11.5cm]{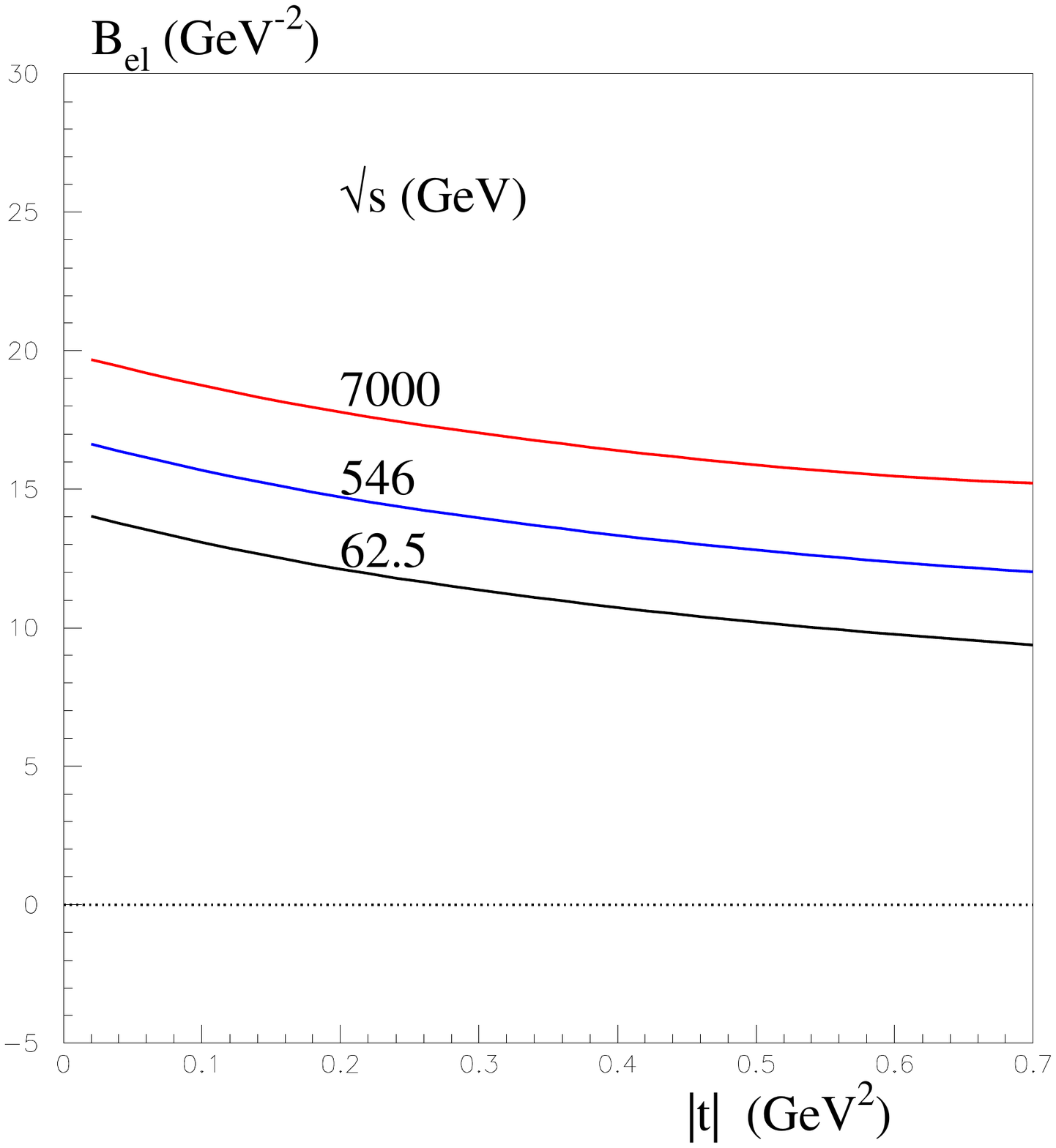}
\caption{\sf The differential proton-proton elastic cross section (left) and the local slope $B$ (right) described by the one pomeron amplitude with the coupling given by the proton form fasctor $F_1(t)$, but without pion loop in the pomeron trajectory; the data are from~\cite{ISR}-\cite{TOTEM}.The parameters are $\Delta=0.08,\ \sigma_0=23$ mb and $\alpha'_P=0.3 $ GeV$^{-2}$.} 
\label{fig:3}
\end{center}
\end{figure}

\begin{figure} 
\begin{center}
\vspace{-3.5cm}
\includegraphics[height=11.5cm]{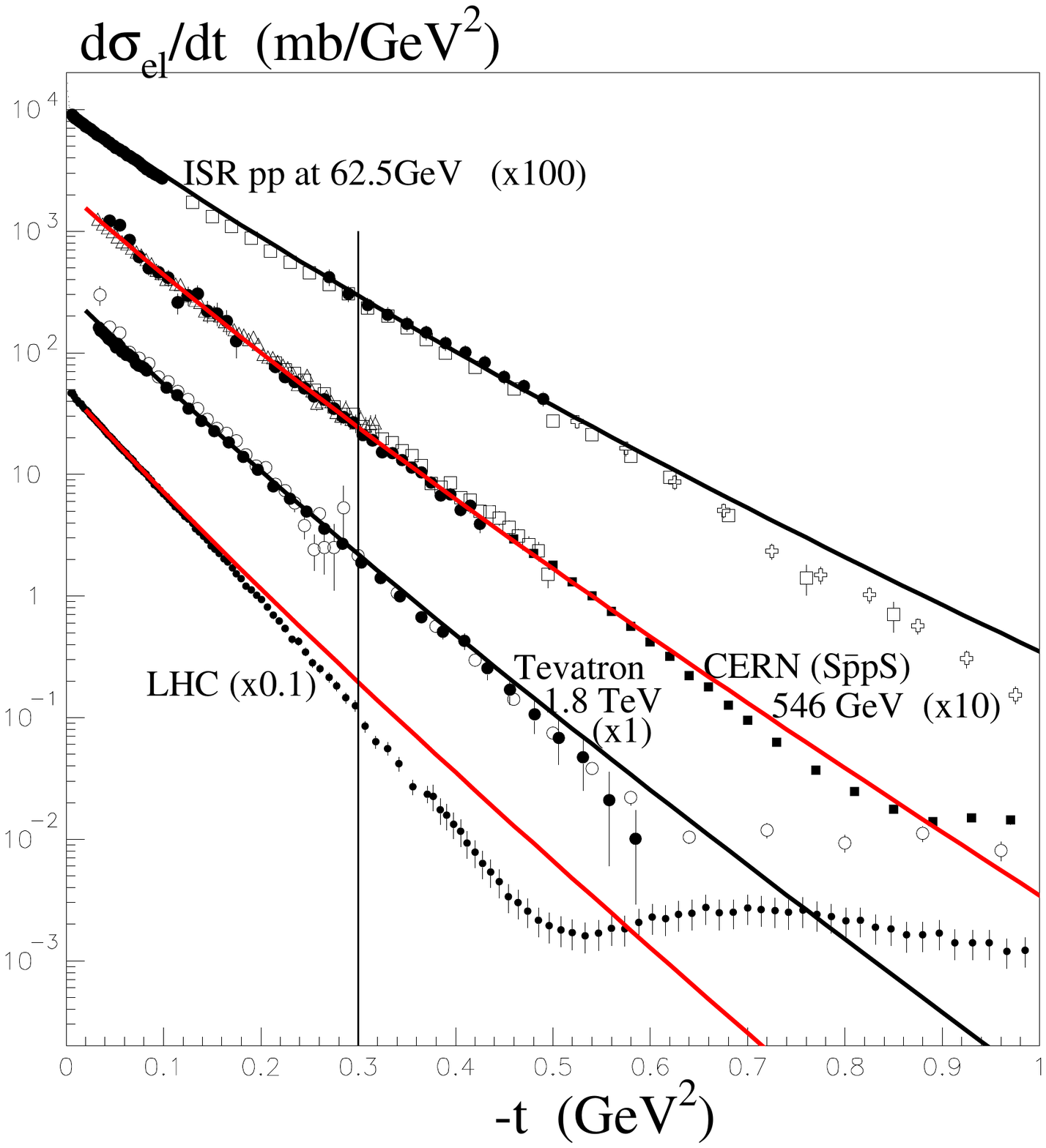}
\includegraphics[height=11.5cm]{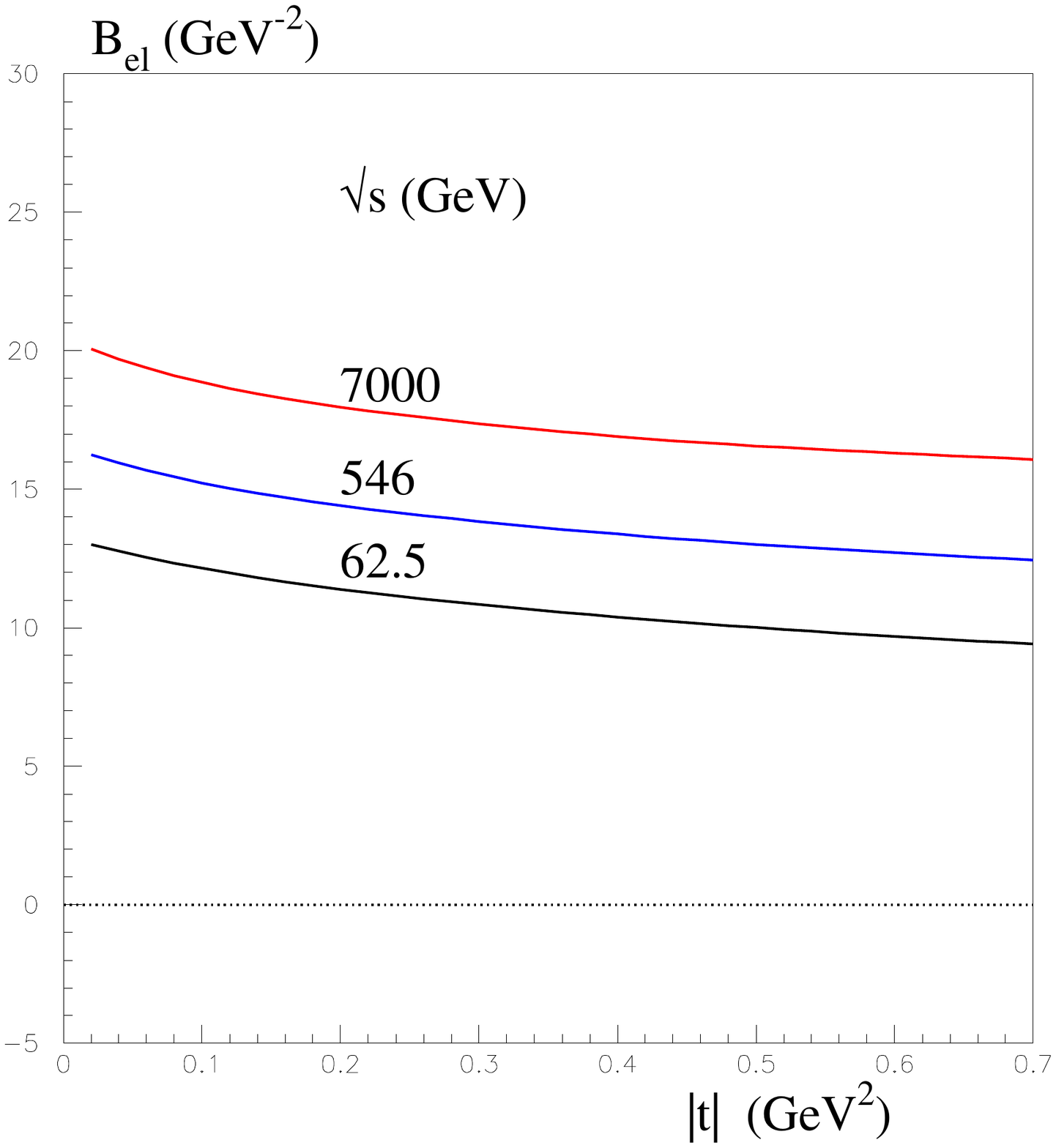}
\caption{\sf The differential proton-proton elastic cross section (left) and the local slope $B$ (right) described by the one pomeron amplitude with the coupling given by the proton form factor $F_1(t)$, and with the pion loop included in the pomeron trajectory; the data are from~\cite{ISR}-\cite{TOTEM}. The parameters are $\Delta=0.085,\ \sigma_0=22$ mb and $\alpha'_P=0.3$ GeV$^{-2}$.} 
\label{fig:4}
\end{center}
\end{figure}

\section{Eikonal rescattering}
The next step is to account for the non-enhanced multi-pomeron diagrams generated by two-particle $s$-channel unitarity
\be
\label{unit}
2~ {\rm Im}T_{\rm el}(s,b)~=~|T_{\rm el}(s,b)|^2+G_{\rm inel}(s,b)\ .
\ee
The unitarity relation (\ref{unit}) is written in the impact parameter, $b$, representation, since, at high energy, the value of $b$ is conserved (to good $\sim 1/s$ accuracy) and plays the role of the orbital angular momentum $l=b\sqrt s/2$.
 $G_{\rm inel}$ is the contribution arising from the sum over all the inelastic intermediate 
 states.

\subsection{One-channel eikonal}
In this case, the solution of the unitarity equation reads
\be
\label{t-omega}
T_{\rm el}(b)~=~i(1-e^{-\Omega(b)/2})\ ,
\ee
where the opacity $\Omega(s,b)$ is described by one-pomeron exchange
\be
\label{opas}
\Omega(s,b)~=~\frac{-i}s\int\frac{d^2q_t}{4\pi^2}~e^{i\vec{q}_t\cdot \vec{b}}~A(s,t=-q^2_t)\ .
\ee
The one-pomeron amplitude $A(s,t)$ is given by (\ref{one}).

\subsubsection{Linear pomeron trajectory and exponential coupling}
The results for the case of a pure exponential proton-pomeron coupling and a linear trajectory (\ref{traj-l}) are shown in Fig.5.  We see that eikonal rescattering 
leads to a strong increase of the local slope $B$ with $|t|$ up to the first diffractive dip (contrary to the effects discussed in the previous section). This is caused by the fact that the absorptive corrections given by the higher $\Omega$ terms of (\ref{t-omega}) have a flatter $t$ behaviour but a negative sign in comparison with one-pomeron exchange.

\begin{figure} 
\begin{center}
\vspace{-3.5cm}
\includegraphics[height=11.5cm]{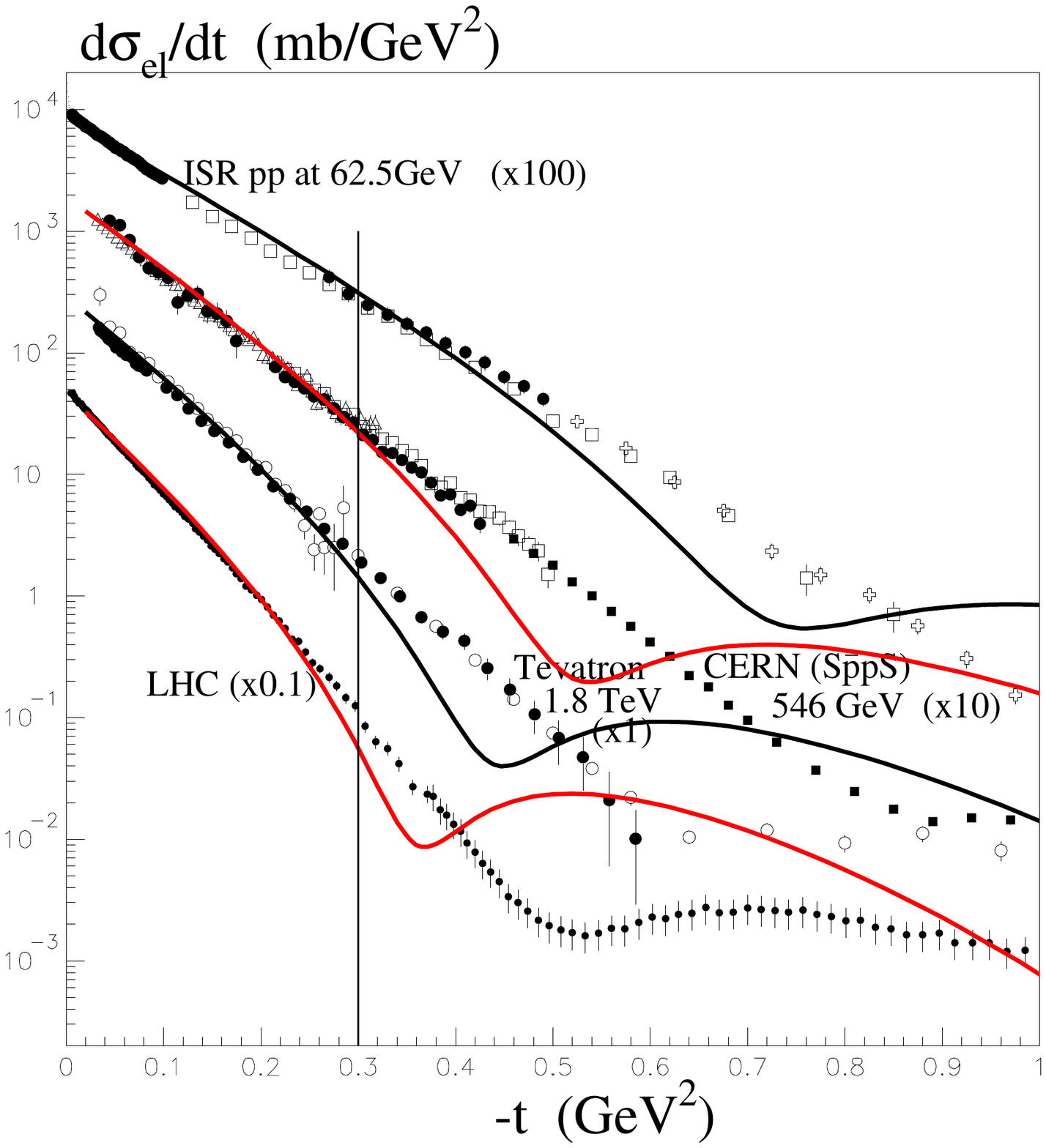}
\includegraphics[height=11.5cm]{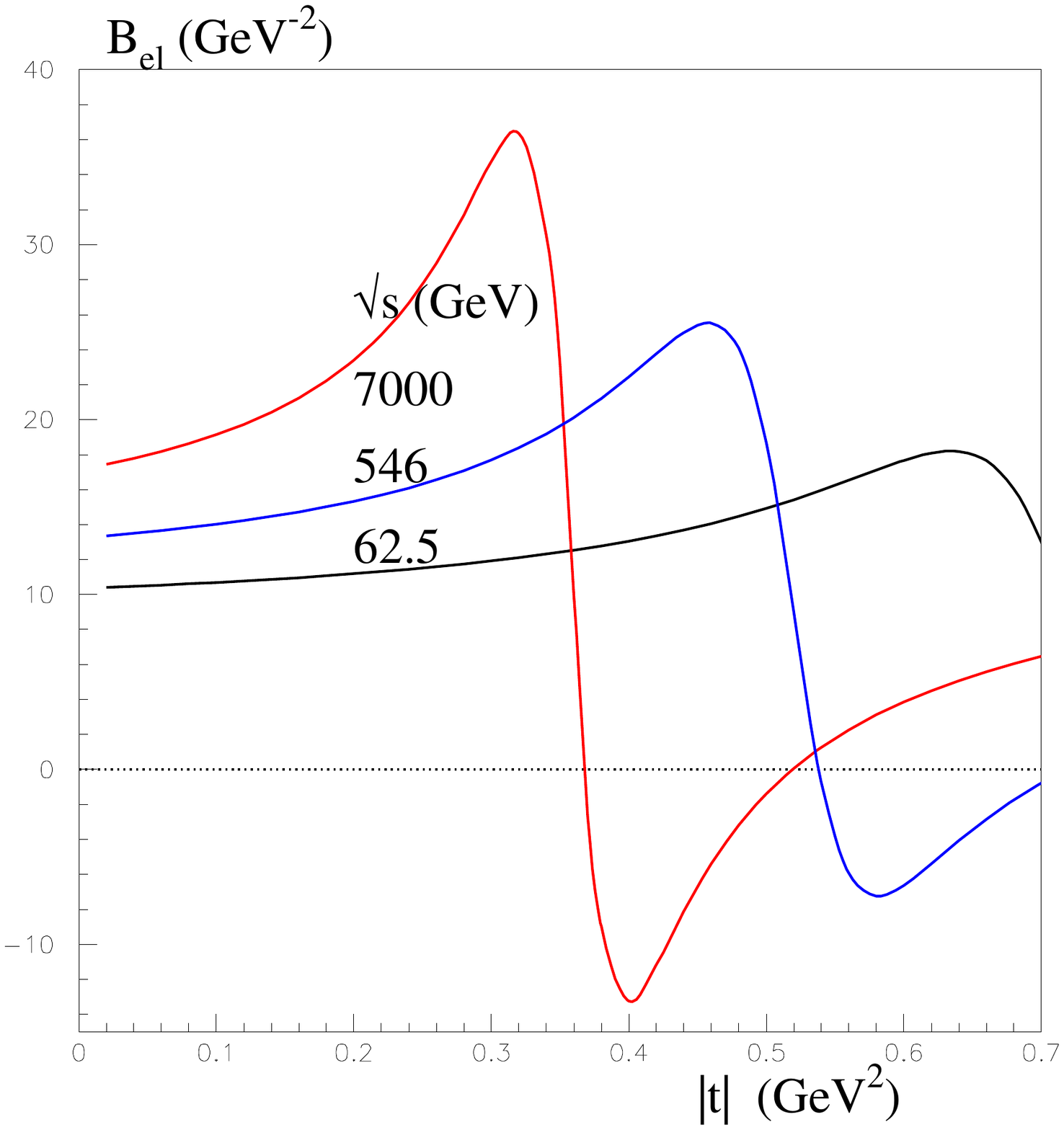}
\caption{\sf The differential proton-proton elastic cross section (left) and the local slope $B$ (right) described by the eikonal amplitude (\ref{t-omega}) with a pure exponential proton-pomeron coupling and with a linear pomeron trajectory; the data are from~\cite{ISR}-\cite{TOTEM}.The parameters are $\Delta=0.11,\ \sigma_0=21$ mb, $b_{exp}=1.2$ GeV$^{-2}$ and $\alpha'_P=0.25 $ GeV$^{-2}$.} 
\label{fig:5}
\end{center}
\end{figure}

\subsubsection{$F_1$ form factor plus the pion loop}
Using a non-exponential coupling $\beta(t)=F_1(t)$ and accounting for the pion loop in pomeron trajectory we get the results shown in Fig.6.
\begin{figure} 
\begin{center}
\vspace{-3.5cm}
\includegraphics[height=11.5cm]{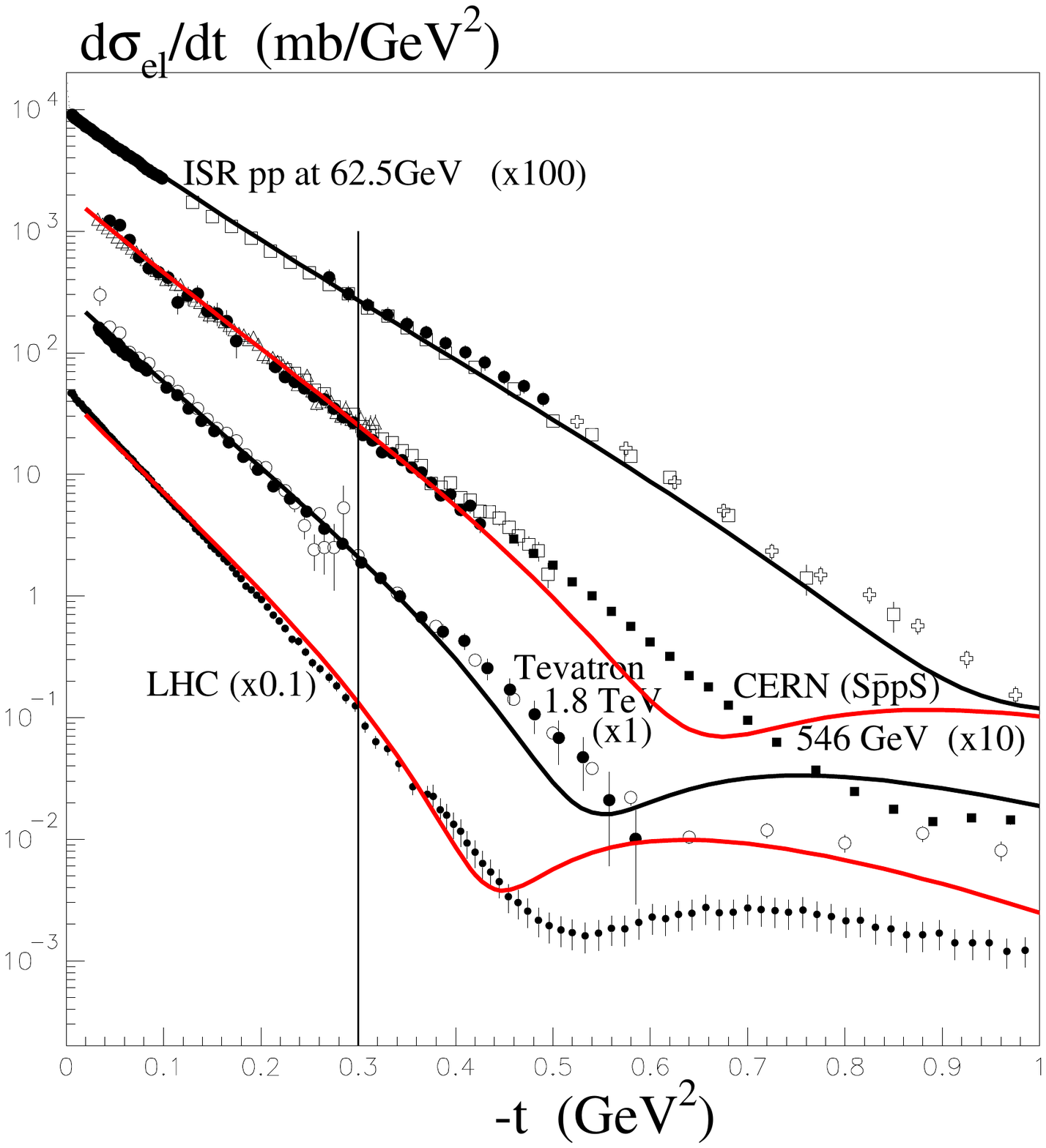}
\includegraphics[height=11.5cm]{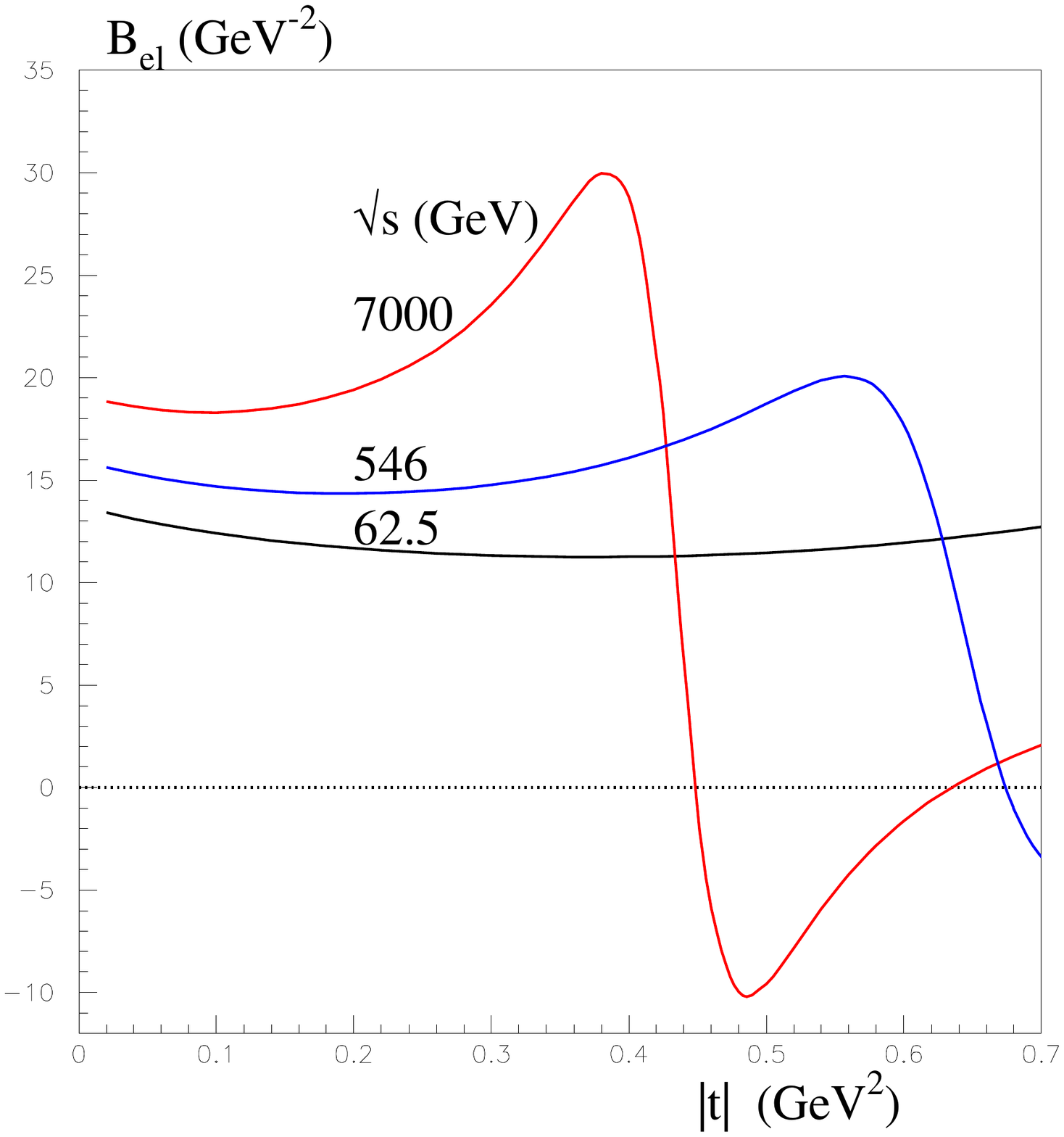}
\caption{\sf The differential proton-proton elastic cross section (left) and the local slope $B$ (right) described by the eikonal amplitude with the coupling given by the proton form factor $F_1(t)$ and with the pion loop included in the pomeron trajectory; the data are from~\cite{ISR}-\cite{TOTEM}. The parameters are $\Delta=0.1,\ \sigma_0=25$ mb and $\alpha'_P=0.12$ GeV$^{-2}$.} 
\label{fig:6}
\end{center}
\end{figure}

\subsection{Two-channel eikonal}
In the models considered above we have not accounted for the possibility of diffractive proton excitation into heavier mass states, such as $p\to N(1440)$, and so on.
It is convenient to include these processes using the Good-Walker formalism~\cite{GW}. This formalism diagonalizes the matrix which describes the
 $p\to N_i$ and $N_i\to N_k$ transitions by introducing the eigenstates $\phi_i$ 
such that the pomeron coupling 
$$\langle\phi_i|A|\phi_k\rangle =A_i\delta_{ik}\, .$$
The proton can then be decomposed into sum of these so-called diffractive eigenstates $\phi_i$, so that
\be \label{ai}
|p\rangle =\sum_i a_i|\phi_i\rangle\, .
\ee
For each state $\phi_i$ the one-channel eikonal formulae of the previous subsection is valid and the elastic scattering amplitude satisfies
\be
\langle p|T|p\rangle ~=~\sum_i|a_i|^2 T_i~=~\langle T\rangle \, .
\ee
The elastic cross section at fixed impact parameter $b$, that is the probability of elastic scattering in a fixed partial wave $l=b\sqrt s/2$, reads
\be
\frac{d\sigma_{el}}{d^2b}~=~\left(\sum_i|a_i|^2T_i \right)^2~=~\langle p|T|p\rangle ^2~=~\langle T\rangle ^2,
\ee
while the probability of diffractive scattering with all possible proton $p\to N^*$ excitations is given by
\be
\frac{d\sigma_{el}}{d^2b}~=~\sum_i|a_i|^2T_i^2~=~\langle p|T^2|p\rangle ~=~\langle T^2\rangle \, .
\ee
Thus the probability of proton (low-mass) dissociation at a given  $b$ is given by the dispersion
\be
\label{SD}
\frac{d\sigma^{\rm SD}}{d^2b}~=~\langle T^2\rangle ~-~\langle T\rangle ^2\, .
\ee

To describe the available collider data, not only on elastic scattering, but including the data on low-mass proton excitations and to get non-zero $\sigma^{\rm SD}_{{\rm low} M}$ we need at least two diffractive eigenstates $\phi_i$. We therefore consider a two-channel eikonal using the Good-Walker formalism. Each eigenstate $\phi_i$ may has its own $i$-pomeron coupling with its own form factor.

The data on diffractive low-mass proton dissociation at collider energies are rather poor. At the relatively low CERN-ISR energy, $\sqrt s=31\ -\ 62$ GeV, the cross section of dissociation of {\em both} protons (that is either of the beam or of the target proton) was evaluated \cite{ISR-SD} to be $\sigma^{\rm SD}_{{\rm low} M}\sim 2\ -\ 3$ mb. At the LHC energy $\sqrt s=7$ TeV the TOTEM result~\cite{T-SD} is $2.6\pm 2.2$ mb, which includes also the probability of dissociation of both protons simultaneously ($\sigma^{\rm DD}$), and is integrated over the mass $M_X<3.4$ GeV  of the outgoing system $X$.
  Besides this there are UA4 data at $\sqrt s=546$ GeV~\cite{UA4SD} --  
 $\sigma^{\rm SD}_{{\rm low} M}=3.0\pm 0.8$ mb for $M<4$ GeV. 
 The non-trivial fact is that 
 this cross section $\sigma^{\rm SD}_{{\rm low} M}$  practically does not appear to increase with energy from the CERN-ISR to the LHC energy regions. On the other hand the elastic cross section, $\sigma_{\rm el}$, increases more than 3.5 times in the same energy interval. At first sight, for one pomeron-driven processes, we would expect both $\sigma_{\rm el}$ and $\sigma^{\rm SD}_{{\rm low} M}$ to have a similar energy behaviour. This point was discussed in~\cite{KMR-E}, so below we will present the local slope calculated within this model. However, first, it is instructive consider a more simpler case.

 \subsubsection{Example of a simple two-channel eikonal model}

In Fig.7 we show the results obtained, within the two-channel eikonal framework, with the coefficients $a_i$ in (\ref{ai}) fixed to be $a^2_1=a^2_2=1/2$, and the $\phi_i$-pomeron couplings taken to be $\sqrt{\sigma_0}\gamma_i\beta_i(t)$ (with $i=1,2$). Moreover, we assume that the $t$ dependence is driven by the same form factor $F_1$ of (\ref{F1}),
\be
\beta_i(t)=F_1(\gamma_it)\, ,
\ee
which means that for each eigenstate $i$, both the value of cross section, $\sigma_0\gamma_i$ and the slope of the couplings are proportional to area (size square) of the component, that is to the value of $\gamma_i$. This, physically reasonable assumption, allows us to decrease the number of free parameters in two-channel eikonal model.
\begin{figure} 
\begin{center}
\vspace{-3.5cm}
\includegraphics[height=11.5cm]{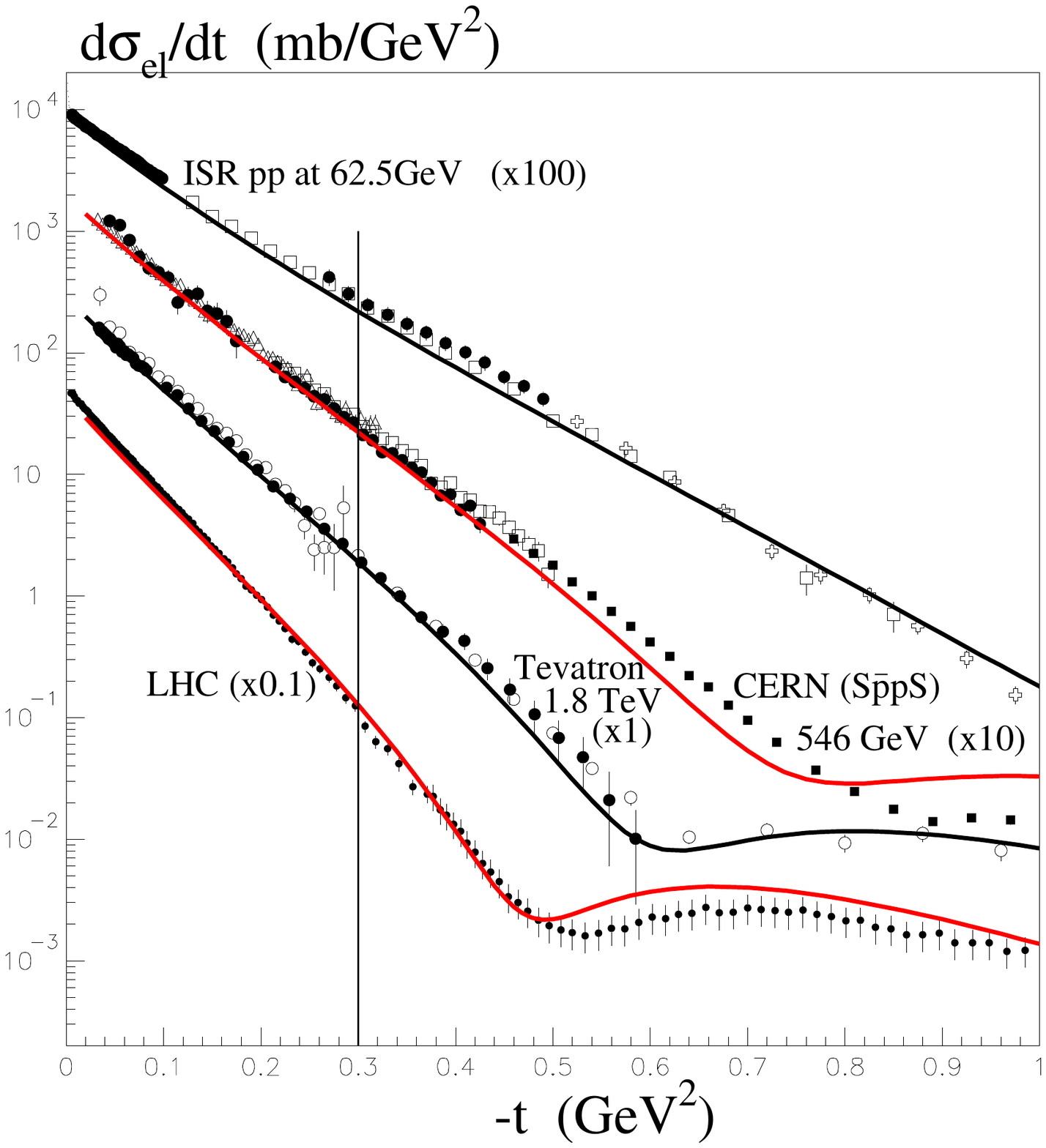}
\includegraphics[height=11.5cm]{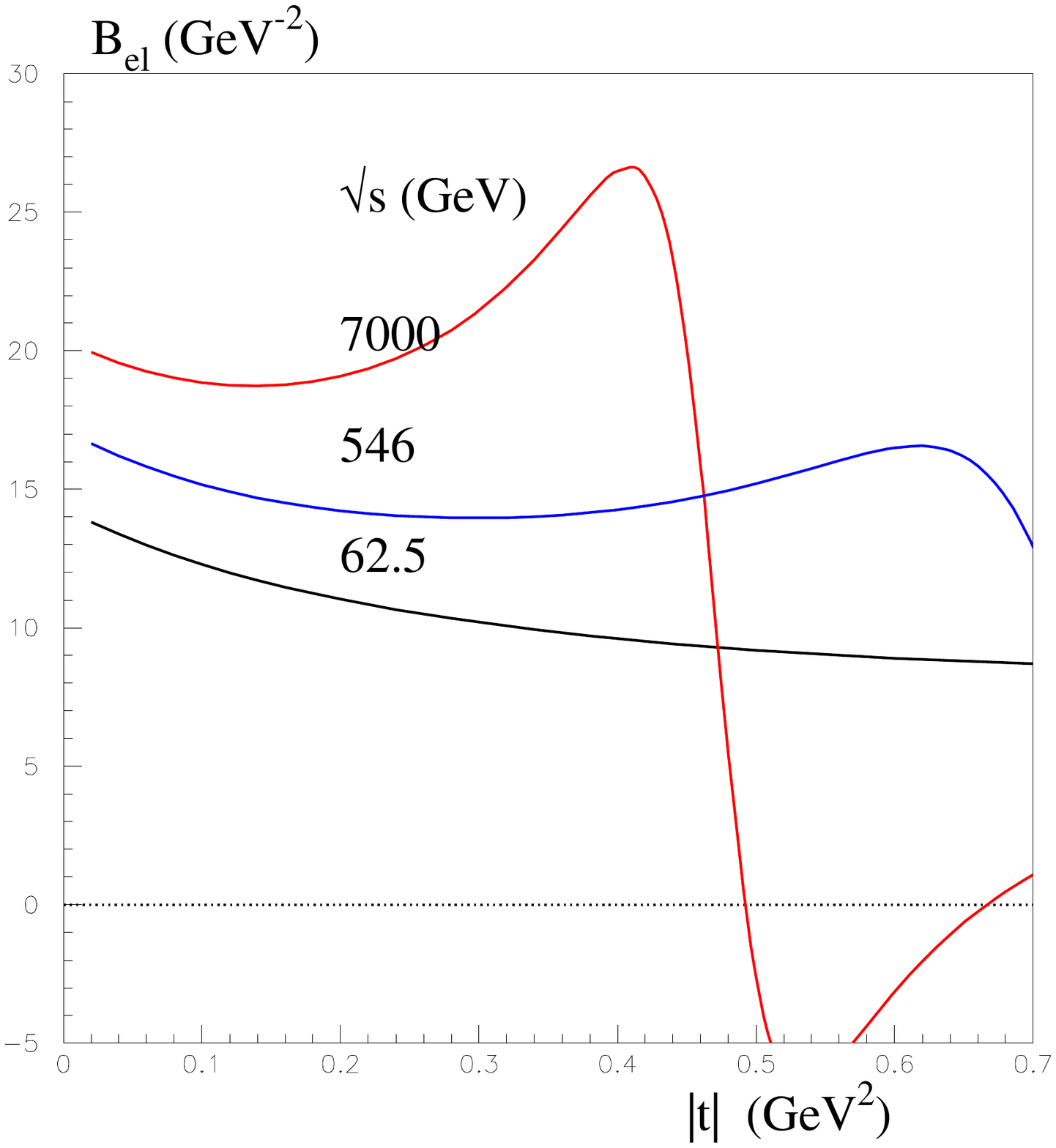}
\caption{\sf The differential proton-proton elastic cross section (left) and the local slope $B$ (right) described by the two-channel eikonal amplitude with the couplings $\sigma_{0,i}=\sigma_0\gamma_i$ and  $\beta_i=F_1(t\gamma_i)$. The pion loop is  included into the pomeron trajectory; the data are from~\cite{ISR}-\cite{TOTEM}. The parameters are $\Delta=0.11,\ \sigma_0=22$ mb and $\alpha'_P=0.1$ GeV$^{-2}$; $\gamma_1=1.38,\,\gamma_2=0.62$.} 
\label{fig:7}
\end{center}
\end{figure}
It is seen in Fig.7 that the {\em two}-channel eikonal allows a better description of the elastic cross sections and that the eikonal induced growth of the local slope at small $|t|$ partly compensates for the decrease of $B$ caused by the non-exponential form of the form factor $F_1$ and the pion loop insertion. 

The probability of low-mass proton excitations in this model is a bit too low at the CERN-ISR energies ($\sigma^{\rm SD}_{{\rm low} M}\simeq 1.9$ mb) and a bit too high at
$\sqrt s=7$ TeV  ($\sigma^{\rm D}_{{\rm low} M}\simeq 4.8$ mb), but within the error bars it does not contradict the data; at $\sqrt s=546$ GeV  the model gives $\sigma^{\rm SD}_{{\rm low} M}\simeq 3.05$ mb.

\subsubsection{Two-channel eikonal tuned to data out to $|t|\simeq0.6~\GeV^2$}
A better description was reached in~\cite{KMR-E} within the two-channel eikonal framework,  but with a larger number of free parameters, now tuned to describe data in an extended $|t|$ domain going beyond the region of the LHC diffractive dip \footnote{The two-channel eikonal framework allows for low-mass dissociation. Besides this, also high-mass dissociation was considered in \cite{KMR-E}. However, it actually does not affect the $t$-slope of the
elastic cross section. The only role of high-mass dissociation here is
a `renormalization' of the effective pomeron trajectory, where the
parameters were anyway tuned to describe the elastic data.}. Interestingly, the form factors of both the diffractive eigenstates turn out to have a behaviour similar to the exp$(-b\sqrt{t})$ used long ago by  Orear et al. \cite{Orear}. The results coming from this version of our model are presented in Fig.8. Recall that in this version the cross sections of low-mass dissociation are in a good agreement with the data: to be explicit the model predicts $\sigma^{\rm SD}_{{\rm low} M}\simeq 2.6$ mb at $\sqrt s=62.5$ GeV, $\sigma^{\rm SD}_{{\rm low} M}\simeq 3.1$ mb at $\sqrt s=546$ GeV and $\sigma^{\rm D}_{{\rm low} M}\simeq 3.75$ mb 
at $\sqrt s=7$ TeV.

 \begin{figure} 
\begin{center}
\vspace{-3.5cm}
\includegraphics[height=11.5cm]{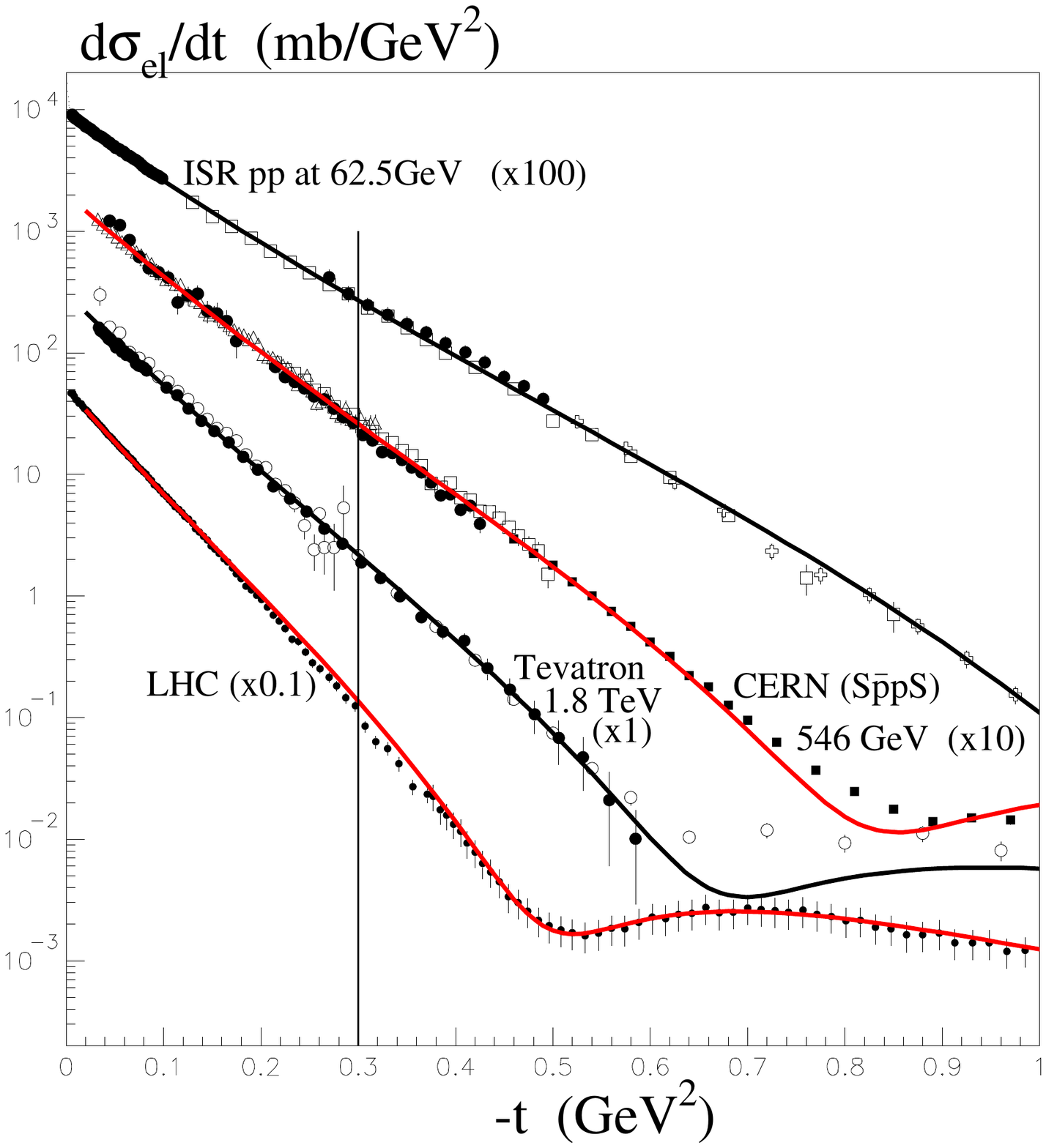}
\includegraphics[height=11.5cm]{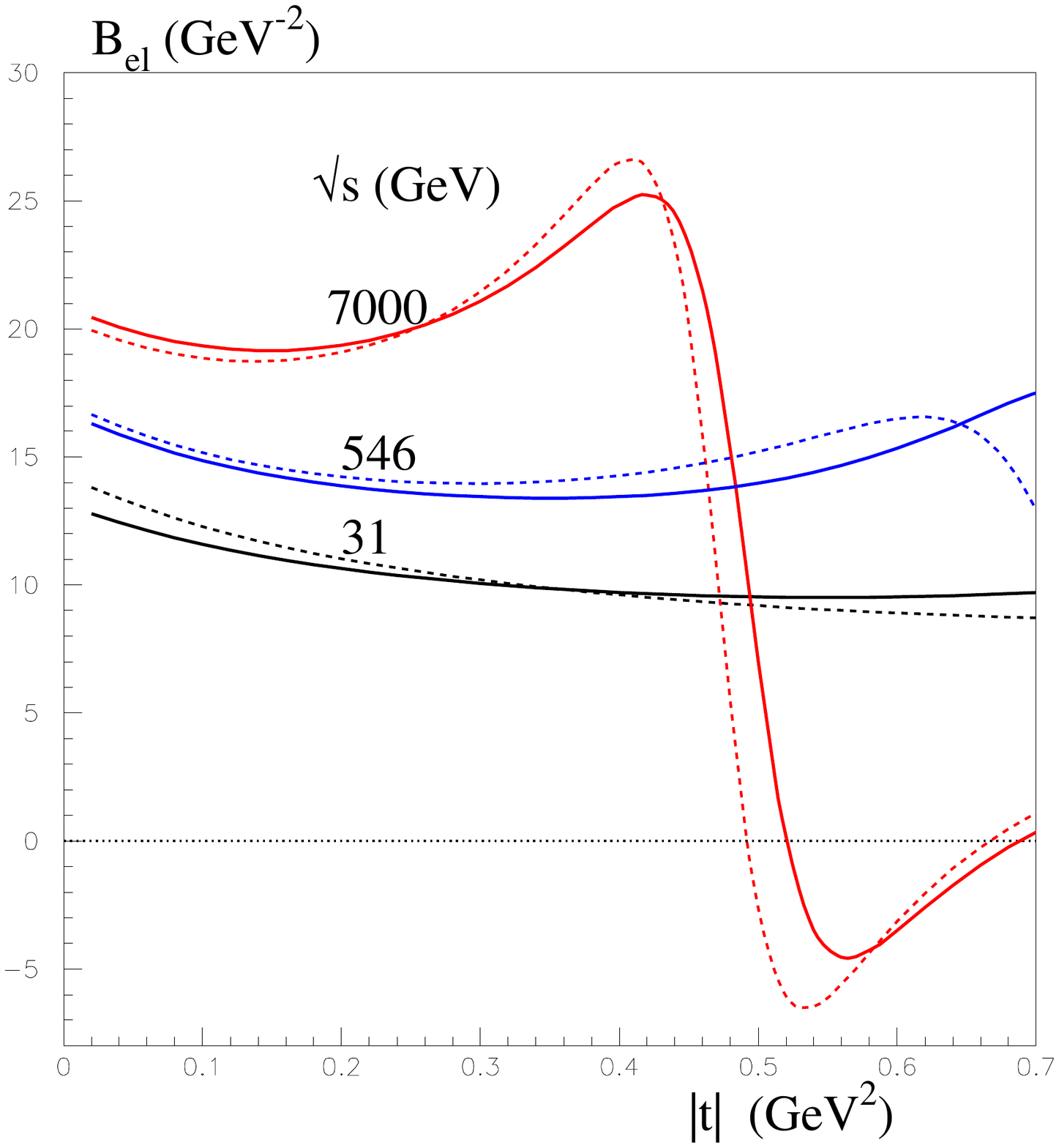}
\caption{\sf The differential proton-proton elastic cross section (left) and the local slope $B$ obtained in the two-channel eikonal model of~\cite{KMR-E} which includes the pion loop contribution to the pomeron trajectory. The local slope from the previous model of Fig.7 is shown by the dashed lines.} 
\label{fig:8}
\end{center}
\end{figure}
For comparison in Fig.8 we show by the dashed lines also the local slope $B(t)$ corresponding to previous toy model of Fig.7.

 Finally we evaluate the expected energy dependence of the local slope $B(t)$ 
 using as example the simplified two channel eikonal described above (see Fig.7). The results are shown in Fig.9.
 \begin{figure} 
\begin{center}
\vspace{-6.5cm}
\includegraphics[height=15.0cm]{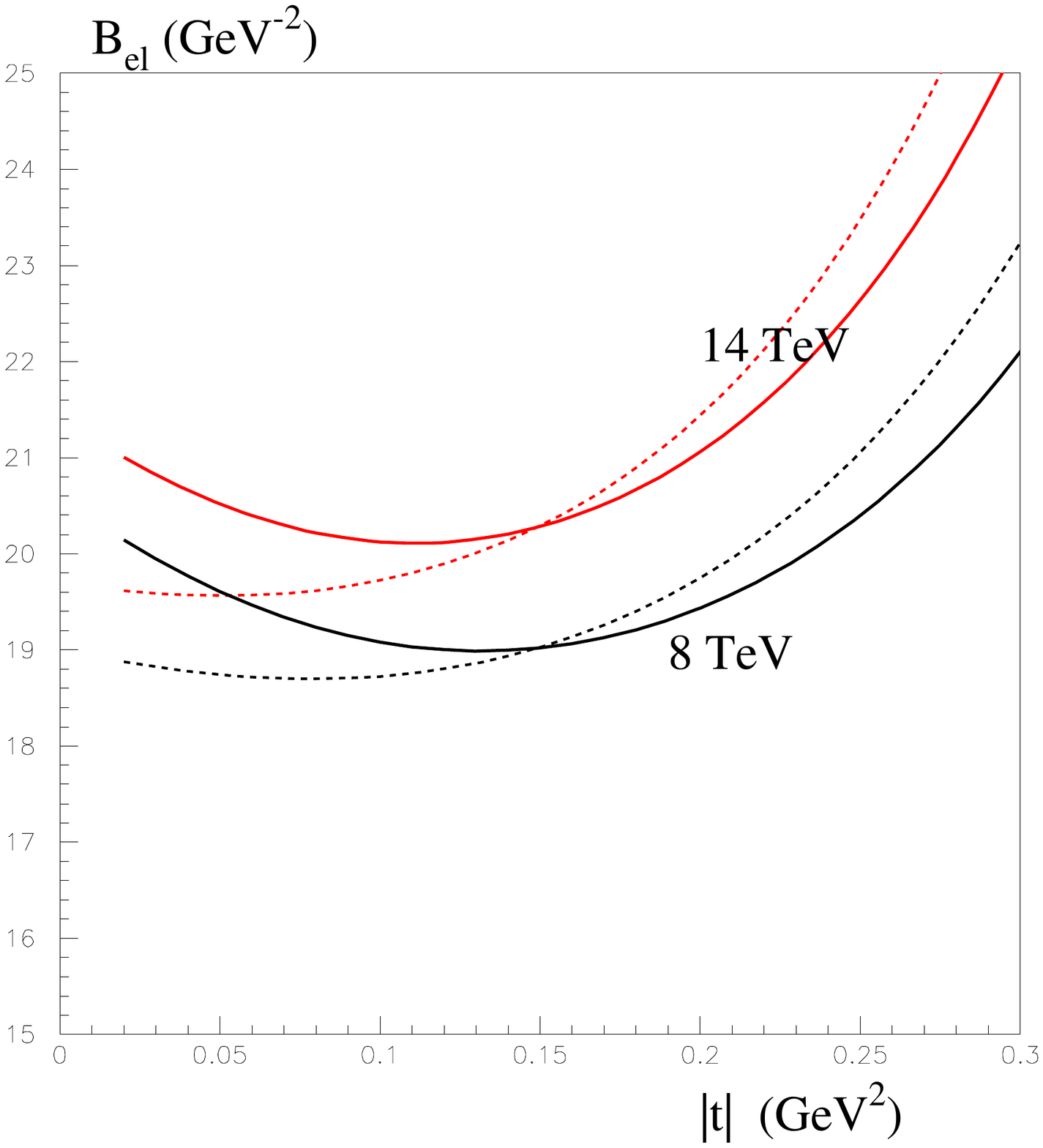}
\caption{\sf  The local slope $B$  calculated at $\sqrt s=8$ and 14 TeV in the two-channel eikonal model corresponding to Fig.7. The local slope in the same model but without the pion loop in the pomeron trajectory is shown by the dashed lines; all other parameters were kept the same
except for the value of the slope of the pomeron trajectory $\alpha'_P$ which was enlarged by $0.04$
GeV$^{-2}$ in order to get the same `mean' slope $B$ and to more or less
satisfactorily describe the data without the pion loop contribution.} 
\label{fig:9}
\end{center}
\end{figure}

.

\section{Discussion}

We have explored the different effects which contribute to the $t$ behaviour of the local slope, $B$, of the proton-proton elastic differential cross section.  We proceeded step-by-step showing how the $t$-behaviours of the different effects lead up to the final overall behaviour shown in Fig.8.  The `pion loop insertion' in the pomeron trajectory and the `pomeron-proton eigenstate couplings' both cause $B(t)$ to decrease with increasing $|t|$, while the `eikonal' effect compensates the decrease resulting in $B(t)$ being surprisingly approximately independent of $t$ out to $-t \simeq 0.3~ \GeV^2$, as required by the data.   Indeed, on the logarithmic
plot the data appear to indicate that $B$ is essentially independent of $t$. However, in fact, a closer inspection reveals that a  characteristic variation with $t$ is expected, as shown in the right-hand plot of Fig.8.  Moreover, the predicted shape of the $t$ behaviour depends on the collider energy, $\sqrt{s}$. 
\begin{figure} 
\begin{center}
\vspace{-6.5cm}
\includegraphics[height=15.0cm]{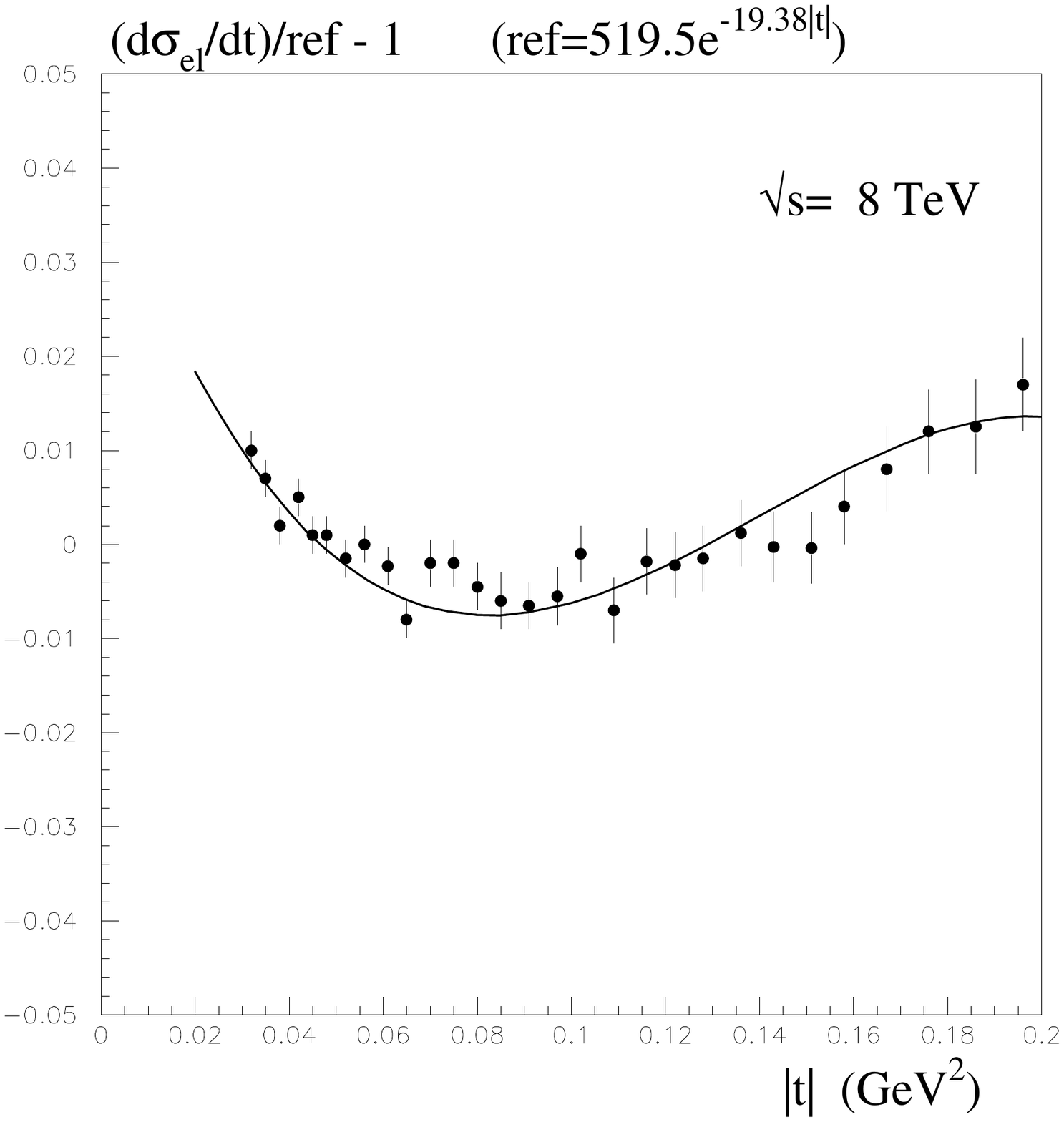}
\caption{\sf The deviation of $d\sigma_{\rm el}/dt$ at 8 TeV from a pure exponential.   The data points are obtained from preliminary measurements by TOTEM and are taken from a presentation at the `Low $x$' meeting \cite{Totem}. The curve corresponds to the model of Fig.8 evaluated at 8 TeV. } 
\label{fig:10}
\end{center}
\end{figure}

We stress that we have primarily been concerned with the behaviour of $B(t)$ at small $t$, namely $-t \lapproxeq 0.3 ~\GeV^2$. Only in Fig.8 have the form factors been tuned to describe the larger $|t|$ data~\footnote{However, as it is seen from Fig.8 (right), this does not change the behaviour of the local slope $B(t)$ in the domain $|t|<0.2$ GeV$^2$ too much.}. Predictions in the region of the diffractive dip require the calculation of the real part of the elastic $pp$ amplitude. This we performed using dispersion relations.

There is already evidence of the expected $t$ dependence at the LHC.  Fig.\ref{fig:10} shows the deviation of $d\sigma_{\rm el}/dt$ from a pure exponential form measured by TOTEM\footnote{The error bars in Fig.\ref{fig:10} do not include systematics which will be discussed in a forthcoming TOTEM publication \cite{TOTEM15}. However, to the best of our knowledge, the qualitative $t$-behaviour is unlikely to be affected by systematics in such a small $t$-interval.} at 8 TeV. These preliminary TOTEM data are compared with the model of Fig.8, recalculated for 8 TeV.
The plot indicates that the expected increase of the $t$-slope $B$  as $t \to 0$ is confirmed \footnote{Throughout this paper we do not consider Coulomb interference effects.}, which hints at, besides the form factor effect, evidence of the pion loop insertion in the pomeron trajectory.

\begin{figure} 
\begin{center}
\vspace{-3.0cm}
\includegraphics[height=11.5cm]{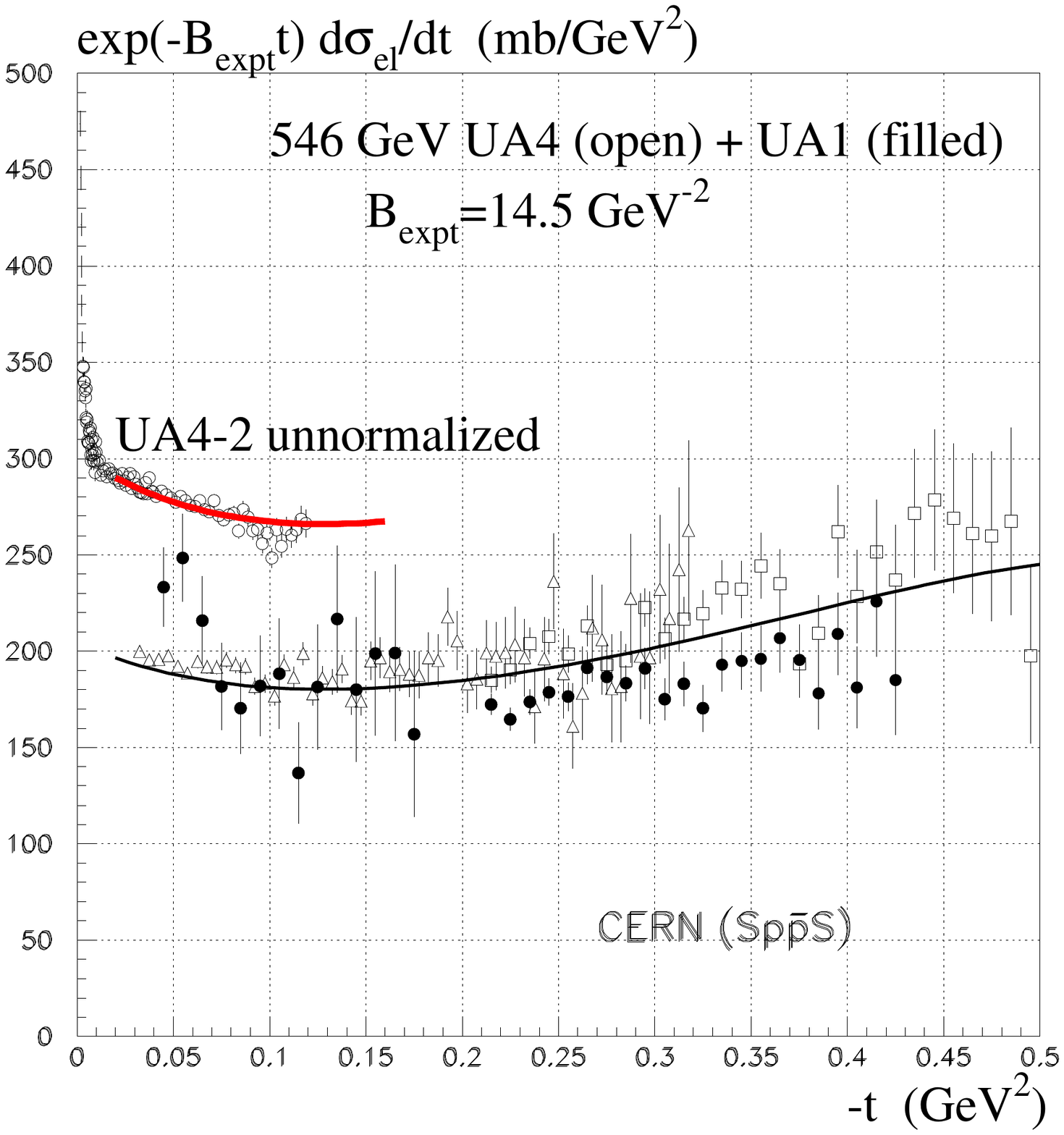}
\includegraphics[height=11.5cm]{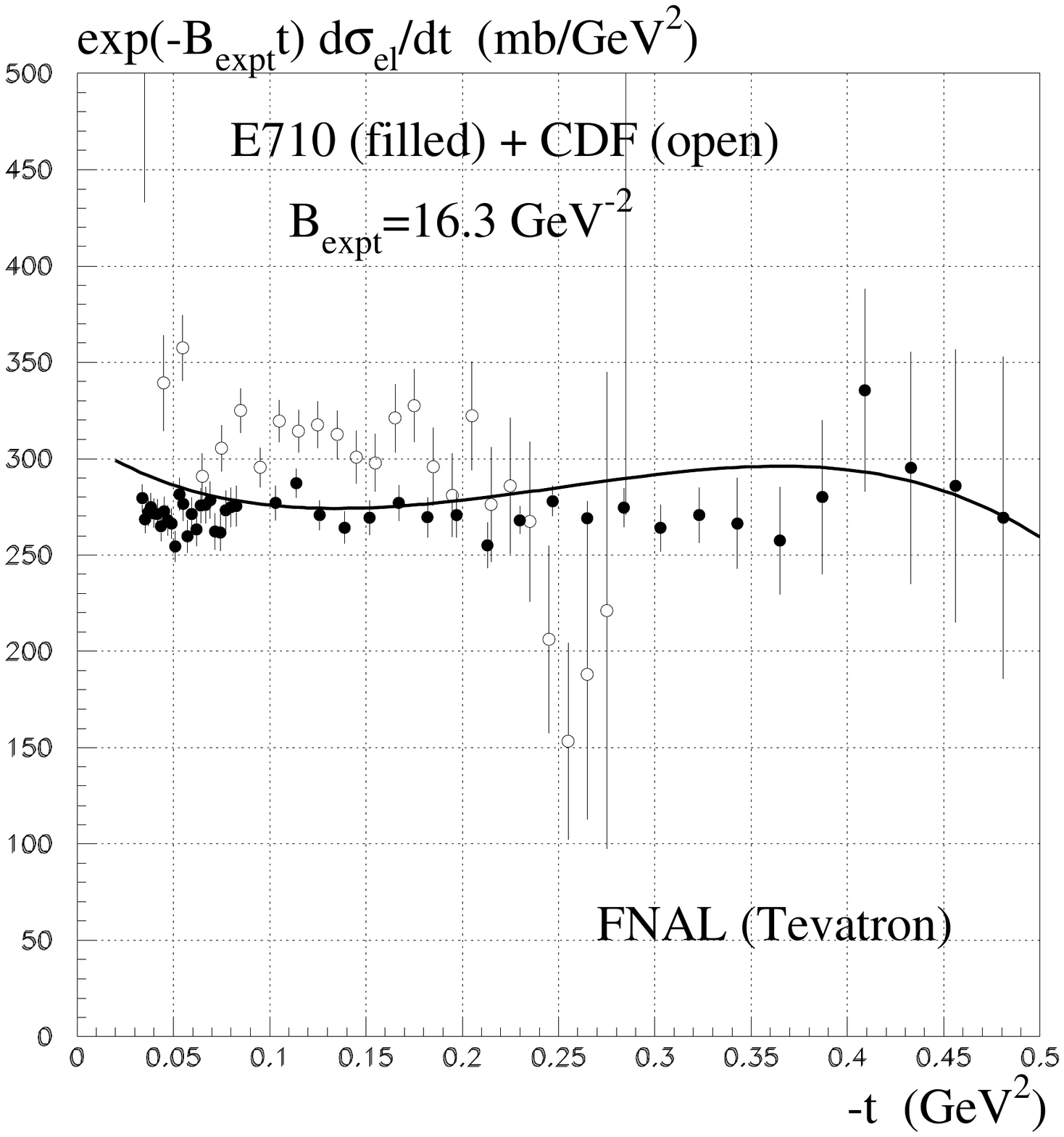}
\caption{\sf The deviation of $d\sigma_{\rm el}/dt$  from a pure exponential form with slope 14.5 and 16.3 GeV$^{-2}$ respectively.   The data points are respectively from~\cite{UA4,UA42,UA1} and \cite{E710,CDFEL} and the curves correspond to the model of Fig.8 evaluated at 546 GeV and 1800 GeV. } 
\label{fig:11}
\end{center}
\end{figure}

The deviations of the differential elastic cross section measured at the collider energies of 546 GeV and 1800 GeV  are shown in Fig.\ref{fig:11}. In comparison with the LHC case shown in Fig.\ref{fig:10} here the curvature produced by the competition of the pion loop and the unitarization (eikonal) effects is much smaller since both effects increase with energy; the pion-loop contribution to the pomeron exchange amplitude is multiplied by $\ln s$ while the screening corrections caused by the eikonalization increase due to the growth of the total cross section.

\section*{Acknowledgements}

MGR thanks the IPPP at the University of Durham for hospitality. This work was supported by the RSCF grant 14-22-00281. We thank Mirko  Berretti and Kenneth  Osterberg for valuable discussions.

\end{document}